\documentclass[a4paper,11pt]{article}
\usepackage{jinstpub} 
\usepackage{lineno}

\title{Wakefield assisted bunch compression in storage rings}

\author{Sergey A. Antipov\note{Corresponding author.},}
\author{Ilya Agapov,}%
\author{Igor Zagorodnov}
\author{and Fran\c{c}ois Lemery}%
\affiliation{%
 Deutsches Elektronen-Synchrotron DESY, Notkestr. 85, 22607 Hamburg, Germany
}%

\emailAdd{Sergey.Antipov@desy.de}

\abstract{Equilibrium bunch lengths typical to modern day light sources of tens of picoseconds limit the spectral reach of superradiant approaches for exploring materials and biological samples. In particular, generation of terahertz radiation seems appealing for pump-probe experiments.
Taking advantage of self-wakes generated by carefully chosen structure one can control the bunch length and shape the bunch profile to increase significantly the high-frequency spectral components. Numerical analysis indicates the presence of self-consistent steady states that can be obtained with passive corrugated or dielectric structures. This approach has potential applications for generating superradiant terahertz radiation and pulse shaping in circular accelerators.}


\begin{document}
\maketitle
\flushbottom

\section{Introduction}\label{sec:intro}

Beyond the production of X-rays in storage rings, terahertz (THz) and mid-infrared (MIR) radiation generation has been widely explored, especially in the context of coherent synchrotron radiation. Early observations indicated that by driving a machine toward the microwave instability (MWI), large broadband and coherent bursting would occur under suitable conditions~\cite{byrd2002,venturini,Roussel}. In parallel, the design of low-momentum-compaction (low-$\alpha$) storage rings were developed to support bunch lengths comparable to the desired radiation wavelength produced in a bending magnet~\cite{BessyTHz}. In addition a low-energy storage ring dedicated to THz generation was proposed~\cite{circe}. Recently feedbacks were employed to maintain a microbunching structure on the beam, allowing the production of stable coherent radiation~\cite{natureSoleil}. THz radiation generation at an energy recovery linac has also been investigated; the smaller achievable bunch lengths in such a platform are favorable to reach larger frequencies and enable the production of large average powers~\cite{Carr2002}.


These techniques all rely on bending magnets e.g. synchrotron radiation to produce THz/MIR, and are generally broadband. Alternatively in the linear accelerator community, there has been tremendous interest in Cherenkov waveguides, e.g. dielectric-lined (DLW), corrugated, or bimetallic waveguides for beam manipulation diagnostics and radiation generation. Beam manipulation studies have especially shown and demonstrated techniques to microbunch beams or to more generally synthesize longitudinal phase spaces ~\cite{antipovTHz, lemeryPRAB, AndonianRamp, LemeryPRL, MayetPRAB}. These effects can be used to enhance the bunch form factor to couple more strongly to similar waveguides which can in turn radiate more energy. Large field gradients (GV/m) have been demonstrated in the superradiant regime where the form factor is approximately unity~\cite{OShea2016}. For a comprehensive discussion on superradiant radiation generation in Cherenkov structures, see ref.~\cite{Floettmann:2020mxv}. We note that a recent technique to characterize such waveguides has been developed~\cite{kellermeier}.

In this paper we investigate the use of Cherenkov waveguides in storage rings. While it is well known that introducing impedances can modify the equilibrium bunch length \cite{haissinski,shobuda,Thomas}, here we further study this feature to longitudinally squeeze the equilibrium bunch length by adiabatically increasing the wakefield amplitude in a corrugated planar structure. The scheme is investigated with analytical and numerical considerations and backed with particle tracking simulations. We demonstrate how the equilibrium bunch distribution can be modified in a controlled way by means of a compact Cherenkov waveguide structure and also discuss limitations of this approach with regards to beam stability. Finally, we also investigate using a secondary wakefield structure to produce radiation with the new equilibrium bunch length.

\section{\label{sec:theory}Bunching for single-mode wakes}
\subsection{Stationary distribution for a single-mode wake}

Consider a relativistic electron bunch orbiting in a storage ring. Let the number of particles in a bunch be $N_b$, the revolution frequency $f_0$, and $\gamma$ the Lorentz factor. Let us further assume a quadratic rf potential well, in which the particles oscillate with an angular synchrotron frequency $\omega_s$. Below the microwave instability threshold the normalized stationary longitudinal bunch distribution $\rho$ in the presence of a wakefield $W^{\prime}$ is described by the Haissinski integral equation~\cite{haissinski}:

\begin{equation}
 \begin{split}
   \frac{\rho(\xi)}{\rho_0(\xi)} & \propto \exp\left [ \kappa
  \int_0^\xi \int_{\xi^{\prime\prime}}^{\infty} \rho(\xi^{\prime}) w^{\prime}(\xi^{\prime\prime} - \xi^{\prime}) d\xi^{\prime}d\xi^{\prime\prime}\right ],\\
  \rho_0(\xi) & = \frac{1}{\sqrt{2\pi}} \exp\left [ -\frac{\xi^2}{2} \right ],
 \end{split}
 \label{eq:one}
\end{equation}
where $\xi = z/\sigma_{z0}$ stands for the normalized coordinate, $\sigma_{z0}$ is the zero intensity rms bunch length, and the normalization condition is $\int \rho(\xi)d\xi = 1$. 
The normalized wake force is $w^{\prime}(\xi) = W^{\prime}(\xi)/W_0$, $W_0 = W^{\prime}(+0)$. The coupling strength is $\kappa = W_0 \frac{\alpha f_0 c}{\gamma \omega_s^2 }\frac{N_b r_0 }{\sigma_{z0}}$; $c$ is the speed of light and $r_0$ the classical electron.

\begin{figure}
 \centering
 \includegraphics[width=.5\linewidth]{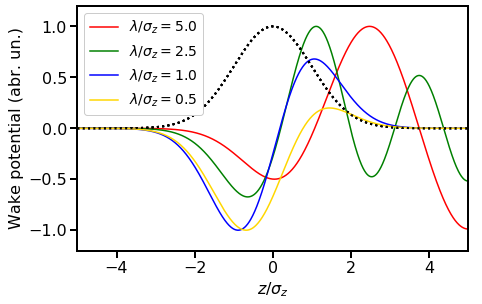}
 \caption{Normalized wake potential for different ratios of the structures frequency to the rms bunch length $
 \lambda/\sigma_z$ for a Gaussian bunch profile (depicted by the dotted black line).}
 \label{fig:X}
\end{figure}

\begin{figure}
 \centering
 \includegraphics[width=.5\linewidth]{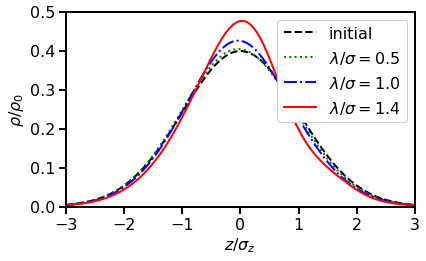}
 \caption{Normalized bunch distributions after compression with a cosine-like wake for different wake wavelengths, found as solutions of Eq.~(\ref{eq:one}). Normalized wake potential $\kappa = 18$. The unperturbed distribution is shown in black.}
 \label{fig:compress_Haisinski}
\end{figure}

Eq.~(\ref{eq:one}) can be solved analytically for certain special wakes~\cite{Burov, Burov-Novokhatski, shobuda, Thomas}. One can see that some wakes can create additional longitudinal focusing, reducing bunch length.
Longitudinal focusing has been shown, for example, for a purely inductive wake~\cite{Thomas}, a purely capacitative wake~\cite{Burov-Novokhatski, Bane:1988wu, Bane:206911}, and for a wake of a ferrite-lined waveguide~\cite{Gerasimov}.
In this paper we consider short-range wakes that can be easily generated in the ring with the help of corrugated or dielectric waveguide structures. This approach seems promising as waveguide structures can be readily engineered to have the desired frequency content in their wakefield and are capable of generating relatively large wakefields over a short distance. As an example consider a Gaussian distribution $\rho(z)$ and a cosine wake $W^{\prime}(z) = W_0 \cos(2\pi z /\lambda)$. Then for $\lambda \sim \sigma_z$ the convolution of the wake with the bunch density creates a longitudinal potential well with the dimensions comparable to $\sigma_z$ (Fig.~\ref{fig:X}). The particles in the head lose energy, while the particles in the tail gain, and the total energy loss by the bunch is minimal. For negligibly weak wakes the solution of Eq.~(1) is simply a normal distribution: $\rho(\xi) = \rho_0(\xi) = \frac{1}{\sqrt{2\pi}} \exp\left [ -\frac{\xi^2}{2} \right ]$. As $W^{\prime}$ increases it produces an extra focusing that reduces the bunch length. In order to obtain the final stationary distribution one can start with a normal distribution and solve Eq.~(1) iteratively. Figure~\ref{fig:compress_Haisinski} presents examples of resulting bunch shapes for different wake wavelengths.


\subsection{Stability limitations}

Eq.~(1) only holds below the MWI threshold. Above the threshold both the bunch length and momentum spread increase thereby limiting the maximum bunch compression that can be achieved with a passive structure. For resonator-like wakes\footnote{Note that Eq.~(\ref{eq:MI_th}) does not apply for some special cases, such as a purely capacitive or a purely resistive wake.} the MWI threshold current for a Gaussian bunched beam can be found from~\cite{Metral:2003cq} 
\begin{equation}
 I_{th} = F\times \frac{\alpha (E/e) \delta_{p,FWHM}^2}{|Z/n|},
 \label{eq:MI_th}
\end{equation}
where $E$ and $e$ stand for the beam energy and electron charge respectively, $F$ is a form factor, $\alpha$ is the momentum compaction factor, and $|Z/n|$ is the effective longitudinal impedance. The second part of the product represents the Keil-Schnell-Boussard stability criterion~\cite{Hofmann:1976xs}, which in case of bunched beams has to be computed with local densities of bunch current $I$ and full-width-at-half-maximum relative energy spread $\delta_{p,FWHM}$. For wakes with $\lambda \sim \sigma_z$, $F \approx 1$.

\begin{figure}
 \centering
 \includegraphics[width=.5\linewidth]{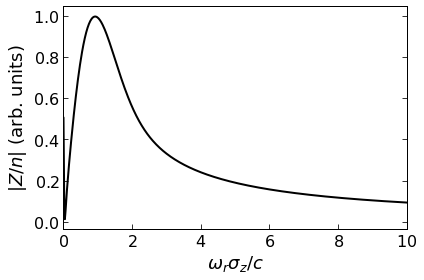}
 \caption{Effective impedance $Z/n$ of a narrow-band resonator as a function of its normalized frequency.}
 \label{fig:Z_over_n}
\end{figure}

The effective impedance is defined as
\begin{equation}
 |Z/n| = \int_{-\infty}^{+\infty} Z(\omega) \frac{\omega_0}{\omega} h(\omega) d\omega,
\end{equation}
where $h(\omega)$ is the normalized beam power spectrum, $\int_{-\infty}^{+\infty} h(\omega) d\omega = 1$ and $\omega_0 = 2 \pi f_0$. For a Gaussian bunch spectrum $h(\omega) \propto \exp(-\omega^2\sigma_z^2/c^2)$ and an impedance peaked at an angular frequency $\omega_r$ the $|Z/n|$ maximizes when $\omega_r \sigma_z/c = 1$ (Fig.~\ref{fig:Z_over_n}).
Thus, increasing the frequency of the impedance seems beneficial for mitigating the instability, potentially achieving larger bunch compression.

It shall be noted as well that in a real machine the instability can be mitigated with other measures, i.e. employing a longitudinal feedback or increasing the momentum spread with an undulator. Thus, Eq.~(\ref{eq:MI_th}) does not necessarily represent a strong limit on the practically achievable bunch compression.

\section{\label{sec:Num_sim}Numerical simulation setup}
In order to assess the effect numerically we performed particle tracking studies using a 1D macroparticle tracking code, where the relativistic electron beam is represented as an ensemble of $N = 10^5$ macroparticles. The code treats the ring as a transfer map with an rf and a wake kick. In order to compute the wake kick the beam is sliced into 100 longitudinal slices, a smooth current density function is computed from the discrete number of particles in each slice using a 2nd order Savitzky–Golay filter, and finally a convolution with the Green's function of the wake is computed.

The initial distribution was assumed to be a 1D Gaussian. We neglected other effects that might also affect the longitudinal distribution in a real machine, like intra-beam scattering, synchrotron radiation damping, or coherent synchrotron radiation. The complete longitudinal wakefield was assumed to be created by the structure alone with no other sources of beam coupling impedance considered. This was done with the sole purpose of simplifying the model and assessing the limits of the proposed approach separately.

\section{\label{sec:Results}Results and Discussion}

In practice there are several ways of generating the desired wake potential in the ring. Here we first consider a fixed aperture dielectric cylindrical waveguide (DLW) that can be treated as a single-mode structure. We also consider a variable gap corrugated structure.

\subsection{\label{sec:Results_single_mode}Single-mode structure}

To gain insights into the beam dynamics and stability of this configuration, we implemented a cylindrically-symmetric DLW with wake potential $W^{\prime}(z) = W_0 \cos(2\pi z /\lambda)$. Here the transverse parameters, e.g. inner radius and dielectric thickness determine the spectral content of the waveguide while the structure length determines the strength of the produced wake~\cite{Rosing:1990gb, Ng}. The wakefields were numerically implemented using publicly available {\sc python} scripts~\cite{PiotPython}. 

\begin{figure}[h]
 \centering
 \includegraphics[width=.5\linewidth]{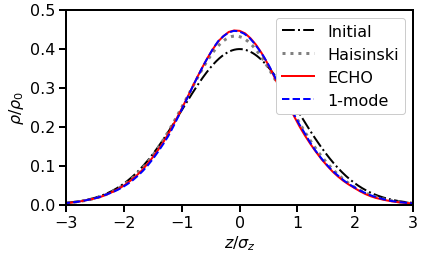}
 \caption{Below the microwave instability threshold the predictions of the Haisinski equation (blue dots) match well the tracking results for a single-mode dielectric structure (orange line) and a corrugated slab structure that creates generates the same wake potential (green line). Wavelength of the 1-mode structure $\lambda/\sigma_z = 0.8$, normalized wake strength $\kappa = 36.12$. The initial distribution is shown in black.}
 \label{fig:Theory_vs_Tracking}
\end{figure}

Considering $\lambda/\sigma_z = 0.8$ and a moderately high normalized wake strength $\kappa = 36.12$ (Example parameters: radius of 5~mm, 7~$\mu$m coating with a dielectric permittivity of 10 for a 400~GHz structure) we observe a measurable bunching of the initial Gaussian distribution in our tracking (Fig.~\ref{fig:Theory_vs_Tracking}). The peak current density increases by $\sim$10\%. This result is in excellent agreement with the prediction of Eq.~(\ref{eq:one}) for the considered wake.

\begin{figure}[h]
 \centering
 \includegraphics[width=.5\linewidth]{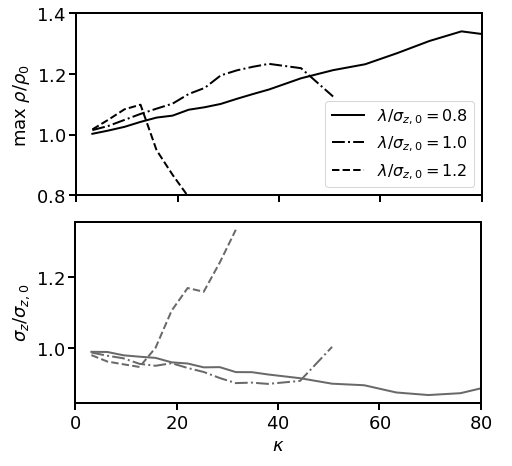}
 \caption{Normalized peak current (a) and bunch length (b) as a function of the normalized strength parameter $\kappa$ for $\lambda/\sigma_{z,0} = 1.2$, $\lambda/\sigma_{z,0} = 1.0$, and $\lambda/\sigma_{z,0} = 0.8$. A shorter wavelength requires a larger wake to achieve the same level of bunch compression, but at the same time allows compressing the bunch further.}
 \label{fig:Peak_Cur_1-mode}
\end{figure}

Larger peak currents can be achieved by increasing the wakefield strength until the MWI threshold. This limit was explored numerically, we first implemented an unperturbed Gaussian distribution and then adiabatically increased $W_0$ in small increments. The peak density increased quasi-linearly until $\kappa \sim 75$, where the bunch started exhibiting unstable behaviour and subsequently rapidly expanded and reduced its peak density; see Fig.~\ref{fig:Peak_Cur_1-mode}. Larger $\lambda$ helps bunching the beam more efficiently, requiring less wakefield strength, but limits the maximum achievable bunch compression to smaller values.

\subsection{\label{sec:Results_slab}Corrugated slab structure}

\begin{figure}[h]
 \centering
 \includegraphics[width=.5\linewidth]{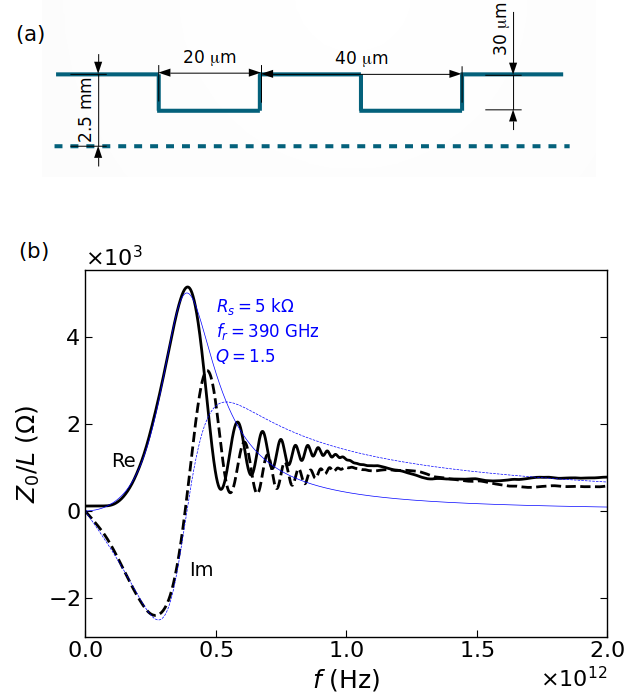}
 \caption{(a) the profile of the corrugated structure (not to scale). Dashed line shows the beam axis. (b) the longitudinal impedance per unit length. Blue lines represent a fit with a single-mode broadband resonator model.}
 \label{fig:Slab}
\end{figure}

In practice it is convenient to generate the required wakefield with a variable-gap slab structure. In addition, slab-structures are advantageous, especially with flat-beams, to mitigate dipole mode excitations which can destabilize the beam ~\cite{slabTremaine}. In such a structure the period and the depth of corrugations mainly govern the resonance frequency while the size of the aperture largely affects the wake strength. Figure~\ref{fig:Slab}a shows an example of such a structure, designed to have a resonance frequency $f_r$ about 400~GHz. The corresponding longitudinal impedance, obtained from ECHO~\cite{Zagorodnov:2017gvp}, is shown in Fig.~\ref{fig:Slab}b; at lower frequencies it can be well approximated with a single-mode broad-band resonator:
\begin{equation}
 Z(f) = \frac{R_s}{1 + i Q (f/f_r - f_r/f)},
\end{equation}
where $R_s$ stand for the resonator shunt impedance and $Q$ for its quality factor. Performing the particle tracking for the same wake strength $\kappa = 36.12$ as in the single-mode case we obtain an identical final beam distribution (Fig.~\ref{fig:Theory_vs_Tracking}). Thus, for a realistic impedance under the MWI threshold the results agree with the theoretical predictions and the simplified single-mode-structure model.

\subsection{\label{sec:Results_example_KARA}KARA ring example}

\begin{table}[t]
\centering
\caption{Key parameters used for the study with a 400~GHz corrugated and a 100~GHz dielectric structures}
\begin{tabular}{l l l} 
 \hline
 Structure & Corrugated & Dielectric \\ [0.5ex] 
 \hline
 Ring circumference & 110~m & 110~m \\
 Bunch energy & 1.3 GeV & 1.3 GeV\\ 
 Bunch charge & $300-700$~pC & $300-700$~pC\\
 Momentum compaction & $5\times 10^{-4}$ & $8\times 10^{-3}$\\
 Bunch length, rms & 3~ps & 10~ps\\
 Synchrotron tune & 0.0023 & 0.005\\
 Structure length & 30~cm & 30~cm\\
 Central frequency & 390~GHz & 100~GHz\\
 Half-gap & 2.5~mm & 5~mm\\
 [1ex] 
 \hline
\end{tabular}
\label{table:1}
\end{table}

As a practical example, we have studied numerically the extent of achievable bunch shortening at a low-$\alpha$ KARA ring at KIT.
The 110-m-long ring is capable of circulating electron bunches of $370-740$~pC and rms bunch length of $2-10$~ps (Table~\ref{table:1}). It has a 40-cm-long straight section suitable for an installation of a passive dielectric or corrugated structure. The $\beta$-function in the straight section and orbit tolerances limit the aperture of the structure to 5~mm. We have considered two cases: a machine operating at $\alpha = 5\times 10^{-4}$ with 3~ps rms bunches and a lattice with a larger momentum compaction of $8\times 10^{-3}$ with 10~ps rms bunches.
\begin{figure}[b]
 \centering
 \includegraphics[width=.5\linewidth]{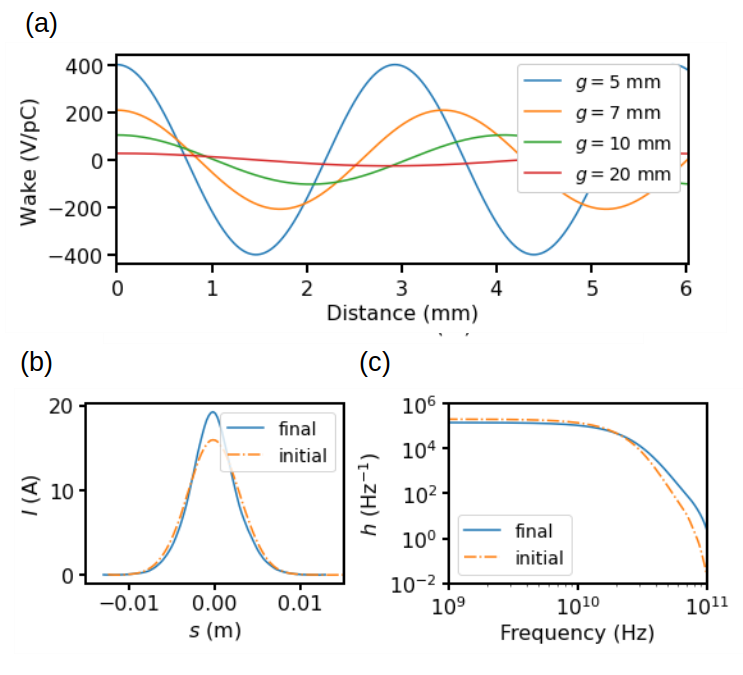}
 \caption{a -- longitudinal wake potential created by a dielectric slab structure vs its half-opening $g$; b -- particle tracking shows how, as the gap of the structure is closed, the electron storage ring the bunches are compressed by 20\%; c -- power spectrum density; the spectral density at the structure's eigenfrequency 100~GHz increases by an order of magnitude. Bunch charge Q = 400~pC.}
 \label{fig:KARA}
\end{figure} 

First, let us consider the later case of $\alpha = 8\times 10^{-3},~ \sigma_{z,0}/c = 10$~ps~(Table~\ref{table:1}). A dielectric structure with the dielectric thickness of 100~$\mu$m and a 5~mm half-gap would produce a longitudinal wake with a characteristic frequency of 100~GHz, or $\lambda/\sigma_z = 1$. Assuming a 30~cm length the wake strength is $W_0 = 4\times 10^{14}$~V/pC, or $\kappa \approx 30$. We simulated numerically the process of slow adiabatic closure of the structure gap of 20~mm, where its wake is negligible, to 5~mm~(Fig.~\ref{fig:KARA}a). As the gap is closed an initially Gaussian bunch transitions to a new equilibrium state with about 20\% smaller rms bunch length~(Fig.~\ref{fig:KARA}b). This greatly boosts the bunch's high frequency spectral components. Figure~~\ref{fig:KARA}c shows a comparison of the initial and the final bunch power spectra: the components close to 100~GHz are enhanced by an order of magnitude. 

\begin{figure}[b]
 \centering
 \includegraphics[width=.5\linewidth]{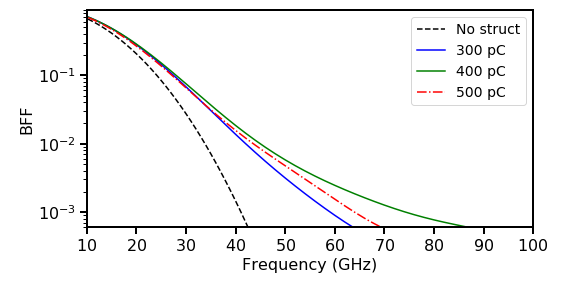}
 \caption{Bunch form factor can be increased at higher frequencies when using the corrugated structure. Different lines correspond to different bunch intensities.}
 \label{fig:KARA-Diel-BFF}
\end{figure}

It is convenient to characterize the temporal structure of the bunch in terms of the bunch form factor (BFF). The BFF is commonly used to characterize the performance of accelerator-based radiation source~\cite{Nodvick:1954zz}. In 1D it is computed as
\begin{equation}
 \widetilde{F}(\omega)^2 = \frac{1}{N^2} \left( \left|\sum_{i = 0}^{N} \cos \frac{\omega z_i}{c}\right|^2 + \left|\sum_{i = 0}^{N} \sin \frac{\omega z_i}{c}\right|^2 \right),
 \label{eq:BFF} 
\end{equation}
where $z_i$ stand for individual positions of the macroparticles. The observed shortening of the bunch greatly increases the BFF at high frequencies. The boost is at least an order of magnitude at the frequencies above 40~GHz (Fig.~\ref{fig:KARA-Diel-BFF}).

A further decrease of the bunch length does not seem to be feasible as a larger wake strength leads to a longitudinal instability. This results in an increase of the bunch length, as seen, for example, by the drop of the BFF at high frequencies when increasing the bunch charge above 400~pC~(Fig.~\ref{fig:KARA-Diel-BFF}).

\begin{figure}[h!]
 \centering
 \includegraphics[width=.99\linewidth]{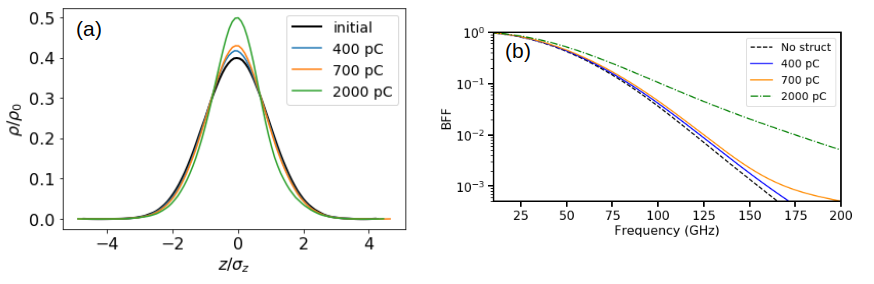}
 \caption{Bunch shortenings that can be obtained with a 30-cm-long 400~GHz corrugated structure in the KARA ring for different bunch charges (a) and the corresponding bunch form factors (b). The initial bunch length is assumed to be 0.9~mm, rms (shown as black lines). }
 \label{fig:Cur_Ampl_Corrugated}
\end{figure}

Accessing even higher frequencies, on the order of 100~GHz, requires shorter bunches. Let us now consider the case of $\alpha = 5\times 10^{-4}$, when the rms bunch length $\sigma_{z,0}/c = 3$~ps. A 30-cm-long corrugated structure can be installed in the same straight section. Considering the profile and impedance shown in Fig.~\ref{fig:Slab} we obtain again a ratio $\lambda/\sigma_z \approx 1$.
At the smallest gap of 5~mm the structure would generate a 5-10\% increase of the peak current for the operational range of bunch charges between 400 and 700~pC~(Fig.~\ref{fig:Cur_Ampl_Corrugated}a). If the length of the structure can be increased or, alternatively, its gap reduced the amplification of the peak current can be even larger (seen in the plot as a hypothetical 2000~pC line). The bunch shortening, as expected, boosts the bunch form factor at higher  frequencies, above 100~GHz (Fig.~\ref{fig:Cur_Ampl_Corrugated}b). 

In all the simulations for the combination of available aperture and structure length we stayed below the MWI threshold. On the other hand, our tracking studies with the single mode structure indicates a potential 30\% increase of peak current before the impedance of the structure starts triggering the MWI (Fig.~\ref{fig:Peak_Cur_1-mode}). If the MWI can be avoided, for example, with the use of a longitudinal feedback or damping wigglers, the bunching could potentially be even greater. 

In comparison, achieving a similar $20-30\%$ increase in the peak current (shown in Fig.~\ref{fig:Peak_Cur_1-mode}) by using a conventional rf-based approach would require increasing the voltage by $40-60\%$, as the bunch length scales as $\sigma_z \propto 1/\sqrt{V_{rf}}$. Alternatively, one can reduce the momentum compaction, $\alpha$, of the lattice. On the other hand, this might compromise the beam stability as the thresholds of the microwave instability and the transverse mode coupling instability are proportional to $\alpha$, while the rms bunch length -- $\sigma_z \propto \sqrt{\alpha}$. Therefore, employing passive structures for bunching might be an appealing option.

\subsection{PETRA~IV ring example} 
As another practical example, let us consider a state-of-the art synchrotron light source PETRA~IV that is currently being planned at DESY~\cite{PETRA_IV_CDR}. This 2304-m-long ring has a beam energy of 6~GeV and is capable of circulating trains of electron bunches with beam currents up to 200~mA. In the brightness operation mode the ring stores trains of 40-ps-long (rms) 1~nC bunches with a bunch spacing of 4~ns (Table~\ref{table:2}). Thanks to its high energy physics heritage, the ring has four long and four short straight sections suitable for installation of additional components. 

\begin{table}[b]
\centering
\caption{PETRA~IV parameters used for the study}
\begin{tabular}{l l} 
 \hline
 Ring circumference & 2304~m\\
 Bunch energy & 6.0 GeV\\ 
 Bunch charge & 400~pC \\
 Momentum compaction & $3.33\times 10^{-5}$\\
 Bunch length, rms & 5~ps\\
 Momentum spread, rms & $9 \times 10^{-4}$\\
 Synchrotron tune & 0.005\\
 Structure length & $25-500$~cm \\
 Central frequency & 160~GHz\\
 Half-gap & 5~mm\\
 [1ex] 
 \hline
\end{tabular}
\label{table:2}
\end{table}

\begin{figure}[h!]
 \centering
 \includegraphics[width=.99\linewidth]{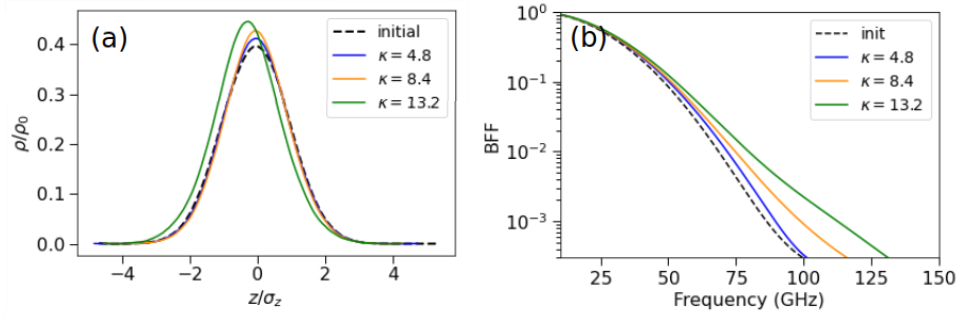}
 \caption{Bunch compression (a) and bunch form factor (b) for different wake strength in PETRA IV.}
 \label{fig:P4_bunching_5ps}
\end{figure} 

The equilibrium bunch length in the ring varies depending on the bunch intensity and the 3rd harmonic cavity voltage. The shortest bunch length that can be achieved in the ring without using its 3rd harmonics system is about 5~ps (1.5~mm) at low beam intensity~\citep{PETRA_IV_CDR}. 

For this example we considered a low intensity bunch of 400~pC and tracked it for 20000 turns (about 100 synchrotron periods), slowly increasing the wake strength. The wake was created by a hypothetical dielectric structure that could be installed in one of the ring's unused undulated cells. The length of the structure was varied from 0 to 5~m (length of the standard undulator), while its half-gap was fixed at 5~mm. We note that a similar wake can also be produced with a corrugated structure (as shown in the previous section). As the wake strength increased (with the 3rd harmonic cavity turned off) the bunch length decreased until eventually the bunch became longitudinally unstable. The maximum bunch compression was achieved for the structure length of 2.25~m, corresponding to the wake strength $\kappa = 13.2$. The maximum bunch compression in this case amounted to a 10\% increase in the peak current density, consistent with the prediction of the compression limit for $\lambda / \sigma_{z,0} = 1.2$ (Fig.~\ref{fig:Peak_Cur_1-mode}). A higher compression can be achieved with a higher frequency structure, which would have to be longer, as discussed in section IV A.  

\section{\label{sec:rad_gen}Radiation generation}
After the bunch length has been reduced to increase the high-frequency spectral components the radiation at the desired frequency can be out-coupled using a separate, dedicated structure. According to~\cite{Floettmann:2020mxv}, the energy radiated at a given frequency $E_{rad}$ is proportional to the BBF:
\begin{equation}
 E_{rad} = N_b^2 e^2 \widetilde{F}^2 L K_{||},
\end{equation}
where $L$ stands for the structure length and $K_{||}$ is the loss factor, i.e. the total energy loss per meter of a charged particle travelling through the structure. Thus, an order of magnitude increase in $\widetilde{F}^2$ results in a significant increase energy $E_{rad}$ radiated at that frequency.

\begin{figure}[h!]
 \centering
 \includegraphics[width=.75\linewidth]{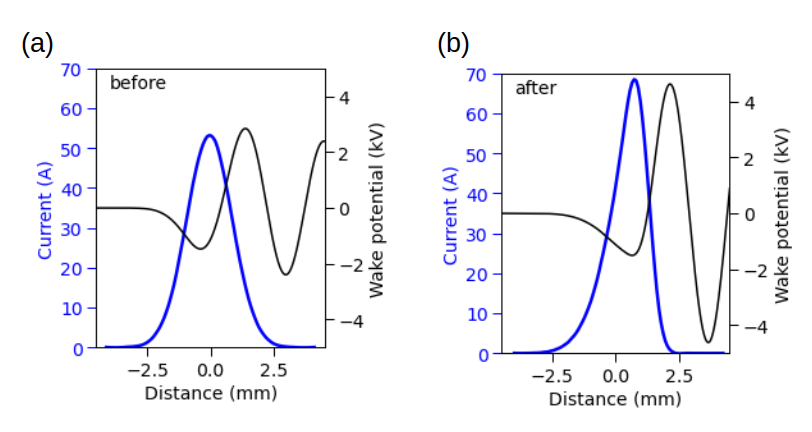}
 \caption{Wake potential of a hypothetical 100~GHz structure for the initial Gaussian bunch (a) and after compression (b) together with the corresponding current profiles. Before compression the energy loss per turn in the structure, defined as the product of the two curves, is negligible and increases dramatically as the bunch is compressed.}
 \label{fig:100_GHz_Rad_Gen}
\end{figure}

To illustrate the benefit of the proposed bunch compression scheme let us consider the hypothetical case depicted by green lines in Fig.~(\ref{fig:Cur_Ampl_Corrugated}): an initially 3~ps rms bunch is compressed until its peak current density increases by 20\%. Let's assume now there is a second structure to out-couple the radiation at 100~GHz. Let it be a 10-cm-long dielectric structure similar to the one considered in section IV C. Initially, a part of the bunch loses the energy to the wakefield, while an almost equal part gains, making the total energy loss, and consequently the radiated power small (Fig.~\ref{fig:100_GHz_Rad_Gen}a). As the current profile changes under the action of the wakefield, its peak density aligns with the minimum of the wake potential and a larger fraction of the beam enters the region of negative wake potential, where particles lose energy (Fig.~\ref{fig:100_GHz_Rad_Gen}b) on each passage. This results in a four-fold increase of the beam energy being emitted in the 100~GHz structure: from $8\times10^{-8}$ to $3\times10^{-7}$~J per passage, or in terms of average radiation power, from 200 to 800~mW. 

For the PETRA~IV case described in section~IV D the increase in the radiated power at 100~GHz would be four-fold: from $3\times10^{-9}$~J to $2.3\times10^{-8}$~J with the same 10-cm-long dielectric structure. The power radiated at 100~GHz could thus reach as much as 1.4~W with the full filling of 1920 bunches, circulating at 130~kHz, providing an effective repetition rate of 250~MHz. In comparison, a dedicated linac to produce THz, TELBE~\cite{TELBEBS}, delivers pulses of $\sim 1~\mu$J at $50$~kHz, or $\sim 0.05$~W.

\section{\label{sec:conslusion}Conclusion and Outlook}

We have demonstrated that a dielectric or a corrugated structure can be used to compress the bunch length in a storage ring in a controllable way which is more efficient than conventional RF-based approaches. To do this one can use a structure with a characteristic wavelength $\lambda \sim \sigma$ that creates a wake potential that produces longitudinal focusing. Numerical simulations performed for the parameters of the KARA ring show that levels of 10-20\% bunch shortening can be realistically achieved before encountering the microwave instability. Similar levels of additional bunch compression could be achieved in 4th generation light sources, operating at low momentum compaction factors, such as the planned PETRA~IV. While the achievable shortening is moderate, this decrease of bunch length can boost the bunch form factor at the frequencies $\sim 0.1$~THz by up to an order of magnitude. This results in a significant increase of radiation power that can be extracted from the beam at those frequencies. Additional use of 3rd harmonic cavities at 1.5 GHz could be explored in the PETRA IV case to further reduce the bunch length in a similar fashion as the proposed CIRCE~\cite{circe}.

The outcoupling of microwave radiation can be done in a fashion similar to the MLS or UVSOR-II setups~\cite{mls,10.1063/1.3463287}, but utilizing the high total beam current of state-of-the art light sources the average THz radiation power can be an order of magnitude larger: 800~mW for our PETRA~IV case vs 60~mW at MLS. While the energy per pulse of $\sim 10~\upmu$J is smaller than $\sim 1$~mJ that can be achieved at an FEL~\cite{Novosib_pump_probe, Zhang:yi5092}, in a storage ring one can obtain a high average power, high repetition rate, broadband source of THz-range radiation. This might be attractive for pump-probe spectroscopy experiments.

The required wakes can be realistically achieved with compact dielectric or corrugated structures. Slab structures particularly are attractive in this context because they can provide control over both the wake amplitude and, partially, the mode frequency. In addition, a more sophisticated corrugated waveguide with  several rows of corrugation placed in parallel would provide a tunable frequency by exposing the beam to the different corrugated patterns. This could be appealing for generating short wavelength microwave radiation in synchrotron light sources. We also note that a reduction of bunch lengths can be contemplated for luminosity improvement of future and existing colliders.

\acknowledgments

We are grateful to the KARA team, in particular to Michael Nasse, Miriam Brosi, Markus Schwarz, for multiple discussions of ring parameters that were beneficial for producing realistic calculations of bunch shortening.


\bibliographystyle{JHEP}
\bibliography{biblio.bib}

\providecommand{\href}[2]{#2}\begingroup\raggedright\begin{thebibliography}{10}

\bibitem{byrd2002}
J.M.~Byrd, W.P.~Leemans, A.~Loftsdottir, B.~Marcelis, M.C.~Martin,
  W.R.~McKinney et~al., \emph{{Observation of broadband self-amplified
  spontaneous coherent Terahertz synchrotron radiation in a storage ring}},
  \href{https://doi.org/10.1103/PhysRevLett.89.224801}{\emph{Phys. Rev. Lett.}
  {\bfseries 89} (2002) 224801}.

\bibitem{venturini}
M.~Venturini and R.~Warnock, \emph{{Bursts of coherent synchrotron radiation in
  electron storage rings : A Dynamical model}},
  \href{https://doi.org/10.1103/PhysRevLett.89.224802}{\emph{Phys. Rev. Lett.}
  {\bfseries 89} (2002) 224802}.

\bibitem{Roussel}
E.~Roussel et~al., \emph{{Microbunching Instability in Relativistic Electron
  Bunches: Direct Observations of the Microstructures Using Ultrafast YBCO
  Detectors}},
  \href{https://doi.org/10.1103/PhysRevLett.113.094801}{\emph{Phys. Rev. Lett.}
  {\bfseries 113} (2014) 094801}.

\bibitem{BessyTHz}
M.~Abo-Bakr, J.~Feikes, K.~Holldack, P.~Kuske, W.B.~Peatman, U.~Schade et~al.,
  \emph{{Brilliant, Coherent Far-Infrared (THz) Synchrotron Radiation}},
  \href{https://doi.org/10.1103/PhysRevLett.90.094801}{\emph{Phys. Rev. Lett.}
  {\bfseries 90} (2003) 094801}.

\bibitem{circe}
J.~Byrd, M.C.~Martin, W.~McKinney, D.~Munson, H.~Nishimura, D.~Robin et~al.,
  \emph{Circe: a dedicated storage ring for coherent thz synchrotron
  radiation},
  \href{https://doi.org/https://doi.org/10.1016/j.infrared.2004.01.017}{\emph{Infrared
  Physics \& Technology} {\bfseries 45} (2004) 325}.

\bibitem{natureSoleil}
C.~Evain et~al., \emph{{Coherent Terahertz synchrotron radiation mastered by
  controlling the irregular dynamics of relativistic electron bunches}},
  \href{https://doi.org/10.1038/s41567-019-0488-6}{\emph{Nature Phys.}
  {\bfseries 15} (2019) 635}
  [\href{https://arxiv.org/abs/1810.11805}{{\ttfamily 1810.11805}}].

\bibitem{Carr2002}
G.L.~Carr, M.C.~Martin, W.R.~McKinney, K.~Jorda, G.R.~Neil and G.P.~Williams,
  \emph{{High power terahertz radiation from relativistic electrons}},
  \href{https://doi.org/10.1038/nature01175}{\emph{Nature} {\bfseries 420}
  (2002) 153}.

\bibitem{antipovTHz}
S.~Antipov, M.~Babzien, C.~Jing, M.~Fedurin, W.~Gai, A.~Kanareykin et~al.,
  \emph{{Subpicosecond Bunch Train Production for a Tunable mJ Level THz
  Source}}, \href{https://doi.org/10.1103/PhysRevLett.111.134802}{\emph{Phys.
  Rev. Lett.} {\bfseries 111} (2013) 134802}.

\bibitem{lemeryPRAB}
F.~Lemery and P.~Piot, \emph{{Ballistic Bunching of Photoinjected Electron
  Bunches with Dielectric-Lined Waveguides}},
  \href{https://doi.org/10.1103/PhysRevSTAB.17.112804}{\emph{Phys. Rev. ST
  Accel. Beams} {\bfseries 17} (2014) 112804}
  [\href{https://arxiv.org/abs/1407.1005}{{\ttfamily 1407.1005}}].

\bibitem{AndonianRamp}
G.~Andonian, S.~Barber, F.H.~O'Shea, M.~Fedurin, K.~Kusche, C.~Swinson et~al.,
  \emph{{Generation of Ramped Current Profiles in Relativistic Electron Beams
  Using Wakefields in Dielectric Structures}},
  \href{https://doi.org/10.1103/PhysRevLett.118.054802}{\emph{Phys. Rev. Lett.}
  {\bfseries 118} (2017) 054802}.

\bibitem{LemeryPRL}
F.~Lemery et~al., \emph{{Passive Ballistic Microbunching of
  Nonultrarelativistic Electron Bunches Using Electromagnetic Wakefields in
  Dielectric-Lined Waveguides}},
  \href{https://doi.org/10.1103/PhysRevLett.122.044801}{\emph{Phys. Rev. Lett.}
  {\bfseries 122} (2019) 044801}
  [\href{https://arxiv.org/abs/1806.09513}{{\ttfamily 1806.09513}}].

\bibitem{MayetPRAB}
F.~Mayet, R.~Assmann and F.~Lemery, \emph{{Longitudinal phase space synthesis
  with tailored 3D-printable dielectric-lined waveguides}},
  \href{https://doi.org/10.1103/PhysRevAccelBeams.23.121302}{\emph{Phys. Rev.
  Accel. Beams} {\bfseries 23} (2020) 121302}
  [\href{https://arxiv.org/abs/2009.12123}{{\ttfamily 2009.12123}}].

\bibitem{OShea2016}
B.D.~O'Shea et~al., \emph{{Observation of acceleration and deceleration in
  gigaelectron-volt-per-metre gradient dielectric wakefield accelerators}},
  \href{https://doi.org/10.1038/ncomms12763}{\emph{Nature Commun.} {\bfseries
  7} (2016) 12763}.

\bibitem{Floettmann:2020mxv}
K.~Floettmann, F.~Lemery, M.~Dohlus, M.~Marx, M.~Ivanyan and V.~Tsakanov,
  \emph{{Superradiant Cherenkov-Wakefield radiation as THz source for FEL
  facilities}},  \href{https://arxiv.org/abs/2005.05640}{{\ttfamily
  2005.05640}}.

\bibitem{kellermeier}
M.~Kellermeier, F.~Lemery, K.~Floettmann, W.~Hillert and R.~Assmann,
  \emph{{Self-calibration technique for characterization of integrated THz
  waveguides}},
  \href{https://doi.org/10.1103/PhysRevAccelBeams.24.122001}{\emph{Phys. Rev.
  Accel. Beams} {\bfseries 24} (2021) 122001}
  [\href{https://arxiv.org/abs/2104.01401}{{\ttfamily 2104.01401}}].

\bibitem{haissinski}
J.~Haissinski, \emph{{Exact longitudinal equilibrium distribution of stored
  electrons in the presence of self-fields}},
  \href{https://doi.org/10.1007/BF02832640}{\emph{Nuovo Cim. B} {\bfseries 18}
  (1973) 72}.

\bibitem{shobuda}
Y.~Shobuda and K.~Hirata, \emph{{The Existence of a static solution for the
  Haissinski equation with purely inductive wake force}}, {\emph{Part. Accel.}
  {\bfseries 62} (1999) 165}.

\bibitem{Thomas}
C.~Thomas, R.~Bartolini, J.I.M.~Botman, G.~Dattoli, L.~Mezi and M.~Migliorati,
  \emph{An analytical solution for the haissinski equation with purely
  inductive wake fields},
  \href{https://doi.org/10.1209/epl/i2002-00319-x}{\emph{Europhysics Letters}
  {\bfseries 60} (2002) 66}.

\bibitem{Burov}
A.V.~Burov, \emph{{BUNCH LENGTHENING: IS IT INEVITABLE?}}, {\emph{Part. Accel.}
  {\bfseries 28} (1990) 47}.

\bibitem{Burov-Novokhatski}
A.V.~Burov and A.V.~Novokhatsky, \emph{{The Device for bunch selffocusing}},
  in \emph{{4th Advanced ICFA Beam Dynamics Workshop: Collective Effects in
  Short Bunches}}, 2, 1990.

\bibitem{Bane:1988wu}
K.L.~Bane, M.H.R.~Donald, A.~Hofmann, J.M.~Jowett, W.S.~Lockman, P.L.~Morton
  et~al., \emph{{BUNCH LENGTH AND IMPEDANCE MEASUREMENTS IN SPEAR}},
  {\emph{Conf. Proc. C} {\bfseries 880607} (1988) 878}.

\bibitem{Bane:206911}
K.L.~Bane, \emph{{Bunch lengthening in the SLC damping rings}},  Tech. Rep.
  \href{http://cds.cern.ch/record/206911}{SLAC-PUB-5177}, SLAC, Stanford, CA
  (1990).

\bibitem{Gerasimov}
A.~Gerasimov, \emph{{Longitudinal Wakefield Focusing: An Unconventional
  Approach to Reduce the Bunch Length in Tevatron}},  Tech. Rep.
  FERMILAB-FN-62XX, Fermilab, Batavia, Illinois (1994).

\bibitem{Metral:2003cq}
E.~Metral, \emph{{Longitudinal microwave instability in lepton bunches}},
  {\emph{Conf. Proc. C} {\bfseries 030512} (2003) 3047}.

\bibitem{Hofmann:1976xs}
A.~Hofmann, \emph{{Single Beam Collective Phenomena: Longitudinal}}, .

\bibitem{Rosing:1990gb}
M.~Rosing and W.~Gai, \emph{{Longitudinal and Transverse Wake Field Effects in
  Dielectric Structures}},
  \href{https://doi.org/10.1103/PhysRevD.42.1829}{\emph{Phys. Rev. D}
  {\bfseries 42} (1990) 1829}.

\bibitem{Ng}
K.-Y.~Ng, \emph{{Wake Fields in a Dielectric Lined Waveguide}},
  \href{https://doi.org/10.1103/PhysRevD.42.1819}{\emph{Phys. Rev. D}
  {\bfseries 42} (1990) 1819}.

\bibitem{PiotPython}
P.~Piot, J.~Wang and F.~Lemery, ``Diwakecyl.''
  \url{https://github.com/NIUaard/DiWakeCyl}.

\bibitem{slabTremaine}
A.~Tremaine, J.~Rosenzweig and P.~Schoessow, \emph{{Electromagnetic wake fields
  and beam stability in slab-symmetric dielectric structures}},
  \href{https://doi.org/10.1103/PhysRevE.56.7204}{\emph{Phys. Rev. E}
  {\bfseries 56} (1997) 7204}.

\bibitem{Zagorodnov:2017gvp}
I.~Zagorodnov, \emph{{Computation of Electromagnetic Fields Generated by
  Relativistic Beams in Complicated Structures}},  in \emph{{2nd North American
  Particle Accelerator Conference}}, p.~WEA1IO02, 2017,
  \href{https://doi.org/10.18429/JACoW-NAPAC2016-WEA1IO02}{DOI}.

\bibitem{Nodvick:1954zz}
J.S.~Nodvick and D.S.~Saxon, \emph{{Suppression of Coherent Radiation by
  Electrons in a Synchrotron}},
  \href{https://doi.org/10.1103/PhysRev.96.180}{\emph{Phys. Rev.} {\bfseries
  96} (1954) 180}.

\bibitem{PETRA_IV_CDR}
K.~Wittenburg, \emph{PETRA IV: Upgrade of PETRA III to the Ultimate 3D X-ray
  Microscope} (10, 2019),
  \href{https://doi.org/10.3204/PUBDB-2019-03613}{10.3204/PUBDB-2019-03613}.

\bibitem{TELBEBS}
B.~Green, S.~Kovalev, V.~Asgekar, G.~Geloni, U.~Lehnert, T.~Golz et~al.,
  \emph{High-field high-repetition-rate sources for the coherent thz control of
  matter}, \href{https://doi.org/10.1038/srep22256}{\emph{Scientific Reports}
  {\bfseries 6} (2016) 22256}.

\bibitem{mls}
A.~Pohl, A.~Hoehl, R.~Müller, G.~Ulm, M.~Ries, G.~Wüstefeld et~al.,
  \emph{Terahertz pump-probe experiment at the synchrotron light source mls},
  in \emph{2013 38th International Conference on Infrared, Millimeter, and
  Terahertz Waves (IRMMW-THz)}, pp.~1--2, 2013,
  \href{https://doi.org/10.1109/IRMMW-THz.2013.6665602}{DOI}.

\bibitem{10.1063/1.3463287}
S.~Kimura, E.~Nakamura, M.~Hosaka, T.~Takahashi and M.~Katoh, \emph{{Design of
  Terahertz Pump—Photoemission Probe Spectroscopy Beamline at UVSOR‐II}},
  \href{https://doi.org/10.1063/1.3463287}{\emph{AIP Conference Proceedings}
  {\bfseries 1234} (2010) 63}
  [\href{https://arxiv.org/abs/https://pubs.aip.org/aip/acp/article-pdf/1234/1/63/11959734/63\_1\_online.pdf}{{\ttfamily
  https://pubs.aip.org/aip/acp/article-pdf/1234/1/63/11959734/63\_1\_online.pdf}}].

\bibitem{Novosib_pump_probe}
Y.~Choporova, V.~Gerasimov, B.~Knyazev, S.~Sergeev, O.~Shevchenko, R.~Zhukavin
  et~al., \emph{First terahertz-range experiments on pump – probe setup at
  novosibirsk free electron laser},
  \href{https://doi.org/10.1016/j.phpro.2016.11.027}{\emph{Physics Procedia}
  {\bfseries 84} (2016) 152}.

\bibitem{Zhang:yi5092}
Z.~Zhang, A.S.~Fisher, M.C.~Hoffmann, B.~Jacobson, P.S.~Kirchmann, W.-S.~Lee
  et~al., \emph{{A high-power, high-repetition-rate THz source for
  pump{--}probe experiments at Linac Coherent Light Source II}},
  \href{https://doi.org/10.1107/S1600577520005147}{\emph{Journal of Synchrotron
  Radiation} {\bfseries 27} (2020) 890}.

\end{thebibliography}\endgroup


\end{document}